From MPD@wmf.univ.szczecin.plThu Sep 24 15:47:37 1998
Date: Thu, 24 Sep 1998 16:17:13 +0100
From: MPD@wmf.univ.szczecin.pl
To: mpd@star.cpes.susx.ac.uk
Subject: alex4.tex

\documentstyle[manuscript,aps]{revtex}
\begin{document}

\newcommand{\be}{\begin{equation}}
\newcommand{\ee}{\end{equation}}
\newcommand{\bea}{\begin{eqnarray}}
\newcommand{\eea}{\end{eqnarray}}

\title{Perturbative String Dynamics Near the Photon Sphere}
\author{Mariusz P. D\c{a}browski \footnote{E-mail: mpdabfz@uoo.univ.szczecin.pl}\\
         {\it Institute of Physics, University of Szczecin, Wielkopolska 15,
          70-451 Szczecin, Poland}
{\ }\\
Alexander A. Zheltukhin \footnote{E-mail: zheltukhin@kipt.kharkov.ua}\\
         {\it Department of Theoretical Physics, NSC Kharkov Institute of 
          Physics and Technology, Akademicheskaya Str. 1, 310108, Kharkov, 
          Ukraine}}
\date{\today}

\maketitle
\begin{abstract}

String dynamics near the photon sphere in Schwarzschild spacetime is considered on the basis of
a perturbative approach with respect to a rescaled string tension as a
small parameter. The perturbative string solution in the zeroth and
first approximation is presented. The perturbative solution corresponds to a 
small deformation of the photon sphere in Schwarzschild spacetime. 

\end{abstract}

\vspace{3.cm}

PACS number(s): 04.70.Bw, 04.70.Dy, 11.25.-w, 11.25 Mj

\newpage

\section{Introduction}

The classical evolution of strings in curved backgrounds is described by a
complicated system of second-order non-linear coupled partial differential
equations which is integrable only for some special configurations 
\cite{integr}. A vast simplification of the
equations of motion arises when one neglects string tension and considers
the null (tensionless) strings \cite{null}. Their
equations of motion are null geodesic equations of General Relativity 
appended by an additional `stringy' constraint. The exact null string
configurations were studied in Schwarzschild and cosmological spacetimes
recently \cite{mar1,mar2,kar,alex}. In particular in Ref. \cite{mar1}
the solution for a null string moving along the photon sphere of the
Schwarzschild spacetime was presented. 

However, the important physical information about tensile string dynamics can be
obtained from studying the approximate solutions of its equations of
motion as it was proposed in Ref.\cite{vesan}. In 
Ref.\cite{alex1,alex2} it was considered the realisation of a 
perturbative scheme which was based on the assumption of a small value
of a rescaled string tension parameter \footnote{Suggestion to consider
string tension as a small parameter was also made in
Ref.\cite{nico}. However, in contradiction to 
Ref.\cite{alex1,alex2} the authors of Ref.\cite{nico,carlos} 
considered a perturbative scheme with null strings as zeroth 
approximation.}  

The objective of this paper is to apply the expansion scheme as
proposed in \cite{alex2} for studying string dynamics in Schwarzschild
spacetime and compare the results with a qualitative picture given in 
\cite{mar1}. 

\section{Rescaled Tension as a Perturbation}

In this Section we shortly discuss the main points of the perturbation
scheme of Ref.\cite{alex2}. The basic idea is to use for the string
action a generalization of the action given for massless point particle  
Ref.\cite{witten}. We assume a perturbative
parameter $\varepsilon \equiv \gamma/\alpha' \ll 1$ with $\gamma$ a
constant and $\alpha'$ the inverse string tension parameter. It was shown
in Ref.\cite{alex2} that for the case of small $\varepsilon \ll 1$ one can
introduce a macroscopic `slow' worldsheet time parameter 
\be 
T = \varepsilon \tau ,
\ee
where $\tau$ is the proper string time parameter. On the scale $T$ the
string oscillations can be considered as perturbations with respect to
the translational motion of the string points and described in the form
of asymptotic expansion   
\be
\label{coord}
X^{\mu}(T,\sigma) = \varphi^{\mu}(T) + \varepsilon \psi^{\mu}(T, \sigma) + \varepsilon^2 
\chi^{\mu}(T, \sigma) + \ldots   ,
\ee
with $\sigma$ beeing a spacelike worldsheet string coordinate and $\mu,
\nu, \rho, \kappa = 0, 1, 2, 3$.  
After introducing the expansion (\ref{coord}) the perturbative equations
of motion and constraints in the first approximation have the form \cite{alex2} 
\bea
\label{dev}
\left( {\cal{D}}_{T}^{2} - \partial_{\sigma}^{2} \right) \psi^{\mu} + 
R_{\nu\rho\kappa}^{\mu}(\varphi) \varphi_{, T}^{\nu} \varphi_{, T}^{\rho}
\psi^{\kappa} & = & 0   \nonumber,\\
\left( \varphi_{\mu , T} {\cal{D}} _{T} \psi^{\mu} \right) & = & 0  ,\\
\left( \varphi_{\mu , T} \psi^{\mu} \right) & = & 0   \nonumber,
\eea
where ${\cal D}_{T} \psi^{\mu} = \psi_{, T}^{\mu} + \varphi_{, T}^{\nu}
\Gamma_{\nu\kappa}^{\mu}(\varphi) \psi^{\kappa}$, $R_{\nu\rho\kappa}^{\mu}$ is the Riemann
tensor and $(\ldots)_{,T} = \partial/\partial T$. First of Eqs.(\ref{dev}) is of the form of the geodesic deviation
equation with an additional term $\partial_{\sigma}^2 \psi^{\mu}$ describing
the elastic string force. The zeroth order equations for
$\varphi^{\mu}(T)$ are geodesic equations for a massless particle in a
given curved space, i.e.,
\bea
\label{gr}
{\cal{D}}_{T} \varphi_{, T}^{\mu} & = & 0   \nonumber,\\
\left( \varphi_{, T}^{\mu} \varphi_{, T \mu} \right) & = & 0   .
\eea

\section{String dynamics in perturbative approach}

Referring to the results of Ref.\cite{mar1} we want to use the 
perturbative scheme of Section II for the discussion of string dynamics 
in Schwarzschild spacetime with the line element given by  
(M is the Schwarzschild mass with the dimension of (length)$^{-1}$,  
$\hbar = c = 1$ and $G$ is Newton constant with the dimension of 
(length)$^2$) 
\be
\label{met}
ds^2 = \left( 1 - \frac{2GM}{r} \right) dt^2 - \frac{dr^2}{\left( 1 - \frac{2GM}{r}
\right)} - r^2 \left( d\theta^2 - \sin^2{\theta} d\phi^2 \right)  ,
\ee
so the set of spacetime coordinates is $X^{\mu} = (t, r, \theta, \phi)$.
 
Our suggestion is to choose $M^{-2}$ to play the role of the constant 
$\gamma$ in the rescaled tension parameter
\be
\varepsilon = \frac{M^{-2}}{\alpha'}   ,
\ee
and the perturbative approach can  be applied for the case when
$M^{-2} \ll \alpha'$. 

In order to solve the perturbative equations (\ref{dev}) for the first 
approximation 
functions $\psi^{\mu}$ we need the solution for the zeroth approximation
functions $\varphi^{\mu}$. The general solution of the geodesic equations
(\ref{gr}) for a massless particle in Schwarzschild spacetime is
well-known \cite{chandra}. Here we want to apply a particular form of the   
solution of the geodesic motion which describes a massless particle moving on
the photon sphere in Schwarzschild spacetime in the form \cite{mar1} 
\be
\label{zer}
\varphi^0(T) = 3ET , \hspace{0.3cm} \varphi^1(T) = 3GM,
\hspace{0.3cm} \varphi^2(T) = \pm \frac{E\tau}{\sqrt{3}GM} + \theta_{0}   ,
\hspace{0.3cm} \varphi^3(T) = \phi_0 ,
\ee
with $E, \theta_0, \phi_0$ constant.  
It is important that (\ref{zer}) automatically satisfies the constraint
(\ref{gr}). 

Substituing (\ref{zer}) into the equations for the first approximation 
functions (\ref{dev}) and making use of the background metric components
(\ref{met}) we get the equations of motion for $\psi^{\mu}$ 
\bea
\label{eqm}
\psi_{, TT}^0 - \psi_{, \sigma \sigma}^0 + 2 \frac{E}{GM}
\psi_{, T}^1 & = & 0   \nonumber,\\
\psi_{, TT}^1 - \psi_{, \sigma \sigma}^1 + 2 \left[
\frac{E}{9M} \psi_{, T}^0 \mp \frac{E}{\sqrt{3}} \psi_{, T} 
- \frac{E^2}{6G^2M^2} \psi^1 \right] & = & 0   ,\\
\psi_{, TT}^2 - \psi_{, \sigma \sigma}^2 & = & 0   \nonumber,\\
\psi_{, TT}^3 - \psi_{, \sigma \sigma}^3 & = & 0   \nonumber,
\eea
and the constraints are \cite{alex2} 
\bea
\label{con}
\psi_{, \sigma}^0 \mp 3 \sqrt{3} GM \psi_{, \sigma}^2 & = & 0   ,\\
\psi_{, T}^0 \mp 3 \sqrt{3} GM \psi_{, T}^2 & = & 0   \nonumber.
\eea
The constraints (\ref{con}) can be integrated to give only one condition 
\be 
\label{con1}
\psi^0 = \mp 3\sqrt{3}GM \psi^2   ,
\ee
and the set of equations (\ref{eqm}) reads 
\bea
\label{eqm1}
\psi_{, TT}^0 - \psi_{, \sigma \sigma}^0 & = & 0   \nonumber,\\
\psi_{, TT}^1 - \psi_{, \sigma \sigma}^1  
- \frac{E^2}{3G^2M^2} \psi^1 & = & 0   ,\\
\psi_{, TT}^2 - \psi_{, \sigma \sigma}^2 & = & 0   \nonumber,\\
\psi_{, TT}^3 - \psi_{, \sigma \sigma}^3 & = & 0   \nonumber.
\eea

After inspecting Eq.(\ref{con}) we can easily notice that
Eqs.(\ref{eqm1}) result in three two-dimensional wave equations for
$\psi^0$, $\psi^2$ and $\psi^3$ which refer to the perturbations in the
components $t$, $\theta$ and $\phi$ of the metric (\ref{met})
respectively. The solutions for these components are
\bea
\label{four}
\psi^0 & = & \sum_{k=-\infty}^{\infty} \left( \alpha_{k}^{0}
e^{ik(\sigma - T)} + \beta_{k}^{0} e^{-ik(\sigma - T)} \right) 
\nonumber ,\\
\psi^0 & = & \sum_{k=-\infty}^{\infty} \left( \alpha_{k}^{2}
e^{ik(\sigma - T)} + \beta_{k}^{2} e^{-ik(\sigma - T)} \right) ,
\\
\psi^0 & = & \sum_{k=-\infty}^{\infty} \left( \alpha_{k}^{3}
e^{ik(\sigma - T)} + \beta_{k}^{3} e^{-ik(\sigma - T)} \right) \nonumber,
\eea
which describe the small string oscillations with frequencies $k = 1, 2
\ldots$ on the macroscopic scale $T$. The emergence of these
oscillations is a pure consequence of admitting the non-zero (but very weak) 
string tension. These oscillations are the oscillations which take place on the
surface of the photon sphere and they do not lead to any deformations of 
this sphere. On the other hand, the equation for the 
radial correction $\psi^1$ in (\ref{eqm1}) has the following solution 
\be
\label{sol}
\psi^1 = A^1 \sin{\left(\frac{E}{\sqrt{3}GM} \sigma \right)} + 
 B^1 \cos{\left(\frac{E}{\sqrt{3}GM} \sigma \right)}   ,
\ee
with $A^1, B^1$ constant. The solution (\ref{sol}) must subject the 
periodicity condition 
\be
\label{per}
\psi(0) = \psi(2\pi)   .
\ee
As a consequence of (\ref{per}) we get the `quantization' of the parameter
$E$  
\be
E = \sqrt{3} GM n  \hspace{1.cm} (n = 0, \pm 1, \pm 2, \ldots) ,
\ee
and the final solution for $\psi^1$ is 
\be
\label{fin}
\psi^1 = A^1 \sin{n\sigma} + B^1 \cos{n\sigma}  .
\ee

\section{Conclusion}

>From (\ref{fin}) we can conclude that opposite to the case of
oscillations for the components $\psi^0$, $\psi^2$ and $\psi^3$ as
described by (\ref{four}) the solution (\ref{fin}) corresponds to a small
static deformation of the photon sphere $r = 3GM + \varepsilon( A^1
\sin{n\sigma} + B^1 \cos{n\sigma} )$ and $\varepsilon \ll 1$. 

This result confirms the qualitative statement of Ref.\cite{mar1} namely
that it is impossible to have the tensile strings with a constant value
of the radaial coordinate $r$ in Schwarzschild spacetime. Besides, in both 
zeroth and first order approximations of the discussed perturbative
scheme only one constraint exists. We note this is in agreement with the
discussion of Ref.\cite{mar1} concerning the absence of the additional 
integral of motion for tensile strings in Schwarzschild spacetime which
suggests their chaotic behaviour. 

The succesful application of the perturbative scheme of Ref.\cite{alex2} in
this paper to the qualitative discussion of \cite{mar1} for the
particular zeroth order solution (a particle on the photon sphere) in 
Schwarzschild spacetime appears to give   
hope that the different exact solutions for the particle motion in
Schwarzschild spacetime \cite{chandra} can be considered as zeroth order
approximations for the perturbative description of the tensile string 
dynamics in this spacetime.

\section{Acknowledgments}
MPD was supported by the Polish Research Committee (KBN) grant No 2 PO3B
196 10. AAZ acknowledges the hospitality of the Institute of Physics,
University of Szczecin.

\end{document}